\documentclass[final,1p,times,times]{elsarticle}
\usepackage{amssymb}
\usepackage{amsthm}
\usepackage{amsmath}
\usepackage{amssymb}
\usepackage{hyperref}
\usepackage{color}
\usepackage{pifont}
\usepackage{algorithm,algorithmic}
\usepackage{mathrsfs}

\newtheorem{theorem}{Theorem}[section]

\newtheorem{prop}{Proposition}[section]

\newtheorem{cor}{Corollary}[section]
\newtheorem{prob}{Problem}[section]
\newtheorem{remark}{Remark}[section]

\DeclareMathOperator*{\argmin}{argmin}
\DeclareMathOperator*{\argmax}{argmax}
\newcommand{\Res}{\textnormal{Res}}

\newcommand{\Htwo}{\mathcal{H}_2}

\DeclareMathOperator{\diag}{\small{\textsf{diag}}\normalsize}
\newcommand{\trace}{\small{\textsf{trace}}\normalsize}
\newcommand{\domega}{\textit{d}\omega}
\newcommand{\ds}{\textit{ds}}
\newcommand{\nynu}{n_{y}\times{n_{u}}}
\newcommand{\nunu}{n_{u}\times{n_{u}}}
\newcommand{\nyny}{n_{y}\times{n_{y}}}
\global\def\mysum{\mathop{\displaystyle\sum}\limits}
\newcommand{\iter}{\small{\textsf{it}}}

\journal{}

\begin{document}



\begin{frontmatter}

%\title{Elsevier \LaTeX\ template\tnoteref{mytitlenote}}
\title{Optimal $\mathcal{H}_{2}$ model approximation based on multiple input/output delays systems}

\author[label1]{I. Pontes Duff}
\author[label2]{C. Poussot-Vassal}
\author[label2]{C. Seren}
\address[label1]{ISAE SUPAERO \& ONERA}
\address[label2]{ONERA - The French Aerospace Lab, F-31055 Toulouse, France}

\begin{abstract}
In this paper, the $\mathcal{H}_{2}$ optimal approximation of a $n_{y}\times{n_{u}}$ transfer function $\mathbf{G}(s)$ by a finite dimensional system $\hat{\mathbf{H}}_{d}(s)$ including input/output delays, is addressed. The underlying $\mathcal{H}_{2}$ optimality conditions of the approximation problem are firstly derived and established in the case of a poles/residues decomposition. These latter form an extension of the tangential interpolatory conditions, presented in~\cite{gugercin2008h_2,dooren2007} for the delay-free case, which is the main contribution of this paper. Secondly, a two stage algorithm is proposed in order to practically obtain such an approximation.
\end{abstract}




\end{frontmatter}


% % % % % % % % % %
% >> main text << %
% % % % % % % % % %
\section{Introduction} \label{s1}

Model approximation plays a pivotal role in many simulation based optimization, control, analysis procedures. Indeed, due to memory and computational burden limitations working with a reduced order model in place of the original one, potentially large-scale, might be a real advantage. To this aim, most of the results presented in the literature address the linear dynamical systems approximation problem in the delay-free case\footnote{''Delay-free case'' means that the approximation model is a dynamical model without any input/output/state delays.}. More specifically, this problem has been widely studied using either Lyapunov-based methods \cite{Spanos:1992, hyland1985optimal, wilson1970optimum}, interpolation-based algorithm \cite{Luenberger1967,gugercin2008h_2,dooren2007,gugercin2009}, or matching moments approaches \cite{grimme1997krylov,astolfi2010model}, leading to a variety of solutions and applications. Recent surveys are available in \cite{antoulas2001survey, antoulas2005approximation, Opmeer2015model}. The  presence of input/output delays in the approximation model was tackled in~\cite{halevi1996reduced} (exploiting both Lyapunov equations and grammians properties derived in \cite{hyland1985optimal} for the free-delay case). The bottleneck of this approach is that it requires to solve Lyapunov equations which might be costly in the large-scale context. From the moment matching side, \cite{scarciottia2014model} proposed a problem formulation that enables the construction of an approximation which contains very rich delay structure (including state delay), but where the delays and the interpolation points are supposed to be a priori known.  From the Loewner framework side, \cite{pontes2015CDC} and  after \cite{schulze2015data}  generalizes the Loewner framework from \cite{mayo2007framework} to the state delay case enabling data-driven interpolation. However, as for the moment matching case, the delays and the interpolation points are supposed to be a priori known.  

In this paper, the problem of approximating a given large-scale model by a low order one including (a priori unknown) I/O delays using the interpolatory framework, is addressed. An alternative ''poles/residues''-based approach is developed, which enables to reach the $\mathcal{H}_{2}$ optimality conditions, treated as interpolation ones. Then, the main contribution of this paper consists in extending the interpolation results of~\cite{gugercin2008h_2} to the case of approximate models with an extended structure, namely, including non-zero input(s)/output(s) delays. Last but not least, $\mathcal{H}_{2}$ optimality conditions for such cases are also elegantly derived together with a single numerical procedure.

%\paragraph{Paper outline} \label{s1ss3}
The paper is organized as follows: after introducing the notations and the mathematical problem statement in Section \ref{section:notation}, Section~\ref{sec:Pre} recalls some necessary preliminary results related to the computational aspects of the $\mathcal{H}_{2}$ inner product and $\mathcal{H}_2$ norm when the calculations are based on the poles/residues decomposition of a transfer function. Section~\ref{sec:MainResult} establishes the $\mathcal{H}_{2}$ optimality conditions solving the input/output delay dynamical model approximation problem. It also proposes an algorithm which permits to practically compute such an approximation. Section~\ref{sec:applications} details the results obtained after treating an academic example. Conclusions and prospects end this article in Section~\ref{sec:conclusion}.   

\section{Notations and problem statement}\label{section:notation}

\paragraph{Notations} \label{s1ss1}
Let us consider a stable {\bf M}ultiple-{\bf I}nput/{\bf M}ultiple-{\bf O}utput ({\bf MIMO}) linear dynamical system, denoted by $\mathbf{G}$ in the sequel, with $n_{u}$ (\emph{resp.}~$n_{y}$) $\in\mathbb{N}^{\ast}$ input(s) (\emph{resp.}~output(s)), represented by its transfer function $\mathbf{G}(s)\in{\mathbb{C}^{n_{y}\times{n_{u}}}}$. Let $\mathcal{H}_{2}^{\nynu}$ be the Hilbert space of holomorphic functions $\mathbf{F}:~\mathbb{C}\rightarrow{\mathbb{C}^{\nynu}}$ which are analytic in the open right-half plane and for which $\int_{-\infty}^{+\infty}\trace\left(\overline{\mathbf{F}(i\omega)}\mathbf{F}^{T}(i\omega)\right)\domega\hspace{-0.05in}<\hspace{-0.05in}{+\infty}$. For given $\mathbf{G},\mathbf{H}\in{\mathcal{H}_{2}^{\nynu}}$, the associated inner-product reads: 
\begin{equation}
\langle\textbf{G},\textbf{H}\rangle_{\mathcal{H}_{2}} =\frac{1}{2\pi}\int_{-\infty}^{+\infty}\trace\left(\overline{\mathbf{G}(i\omega)}\mathbf{H}^{T}(i\omega)\right)\domega,
\label{eq:in_productH2}
\end{equation}
and the $\mathcal{H}_{2}^{\nynu}$ induced norm can be  explained:
\begin{equation}
\|\mathbf{G}\|_{\mathcal{H}_{2}}=\left(\frac{1}{2\pi}\int_{-\infty}^{+\infty}\hspace{-0.15in}\|\mathbf{G}(i\omega)\|^{2}_{\textsf{F}}\domega\right)^{1/2}\hspace{-0.15in}=\langle\mathbf{G},\mathbf{G}\rangle_{\mathcal{H}_{2}},
\label{eq:normH2}
\end{equation}
where $\|\mathbf{G}\|_{\textsf{F}}^{2}=\langle\mathbf{G},\mathbf{G}\rangle_{\textsf{F}}$ and $\langle\mathbf{G},\mathbf{H}\rangle_{\textsf{F}}=\trace(\overline{\textbf{G}}\textbf{H}^{T})$ are the \textit{Frobenius norm} and \textit{inner-product}, respectively. Dynamical system $\mathbf{H}$ will be said \emph{real} \emph{iff.} $\forall{s}\in\mathbb{C},~\overline{\mathbf{H}(s)}=\mathbf{H}(\overline{s})$. It is noteworthy that if $\mathbf{G}(s),\mathbf{H}(s)\in{\mathcal{H}_{2}^{\nynu}}$ are real, then $ \langle\textbf{G},\textbf{H}\rangle_{\mathcal{H}_{2}}= \langle\textbf{H},\textbf{G}\rangle_{\mathcal{H}_{2}}\in{\mathbb{R}_{+}}$.

Besides, any dynamical matrix $\boldsymbol{\Delta}(s)$ will belong to $\mathcal{H}_{\infty}^{\nynu}$ \emph{iff.} $\sup\{\sigma_{max}(\boldsymbol{\Delta}(i\omega))/\omega\in\mathbb{R}\}<{+\infty}$. $\sigma_{max}(\boldsymbol{\Delta}(i\omega))$ refers to the largest singular value of matrix $\boldsymbol{\Delta}(i\omega)$.

Followingly, let $\hat{\mathbf{H}}_{d}$ be a multiple-input/output delays {\bf MIMO} system \emph{s.t.} $\hat{\mathbf{H}}_{d}(s)\in{\mathcal{H}_{2}^{\nynu}}$ and represented by:
\begin{equation}
%\hat{\mathbf{H}}_{d}:~\left\{\begin{array}{rcl}
%\hat{\mathbf{E}}\mathbf{\dot{\hat{x}}}(t)&=&\hat{\mathbf{A}}\mathbf{\hat{x}}(t)+\displaystyle\sum\limits_{j=1}^{n_{u}}\hat{\mathbf{b}}_{j}u_{j}(t-\hat{\tau}_{j})\\
%\hat{y}_{k}(t-\hat{\gamma}_{k})&=&\hat{\mathbf{c}}_{k}^{T}\mathbf{\hat{x}}(t)
%\end{array}\right.,
%\label{eq:DynamicalSystemInputDelay}
\hat{\mathbf{H}}_{d}:~\left\{\begin{array}{rcl}
\hat{\mathbf{E}}\mathbf{\dot{\hat{x}}}(t)&=&\hat{\mathbf{A}}\mathbf{\hat{x}}(t)+\hat{\mathbf{B}}\boldsymbol{\Delta}_{i}(\mathbf{u}(t))\\
\mathbf{\hat{y}}(t)&=& \boldsymbol{\Delta}_{o}(\hat{\mathbf{C}}\mathbf{\hat{x}}(t))
\end{array}\right.,
\label{eq:DynamicalSystemInputDelay}
\end{equation}
%where $\hat{\mathbf{E}},\hat{\mathbf{A}}\in{\mathbb{R}^{n\times{n}}}$ (with state dimension $n\in\mathbb{N}^{\ast}$), $\forall{j}=1\ldots{n_{u}},~\forall{k}=1\ldots{n_{y}},~\hat{\mathbf{b}}_{j},\hat{\mathbf{c}}_{k}\in{\mathbb{R}^{n\times{1}}}$ and $(\hat{\tau}_{k},\hat{\gamma}_{k})\in{(\mathbb{R}^{+})^{2}}$. The transfer function of the underlying system~\eqref{eq:DynamicalSystemInputDelay} from input $\mathbf{\hat{u}}(t)=[\hat{u}_{1}(t)~\ldots~\hat{u}_{n_{u}}(t)]^{T}$ to output $\mathbf{\hat{y}}(t)=[\hat{y}_{1}(t)~\ldots~\hat{y}_{n_{y}}(t)]^{T}$ vectors is given by:
where $\hat{\mathbf{E}},\hat{\mathbf{A}}\in{\mathbb{R}^{n\times{n}}}$ (with state dimension $n\in\mathbb{N}^{\ast}$), $\hat{\mathbf{B}}\in{\mathbb{R}^{n\times{n_u}}}$, $\hat{\mathbf{C}}\in{\mathbb{R}^{n_y\times{n}}}$ and $\boldsymbol{\Delta}_{i}$ and $\boldsymbol{\Delta}_{o} $ are delay operators. The matrix transfer functions $\hat{\boldsymbol{\Delta}}_{i}(s)$ and $\hat{\boldsymbol{\Delta}}_{o}(s)$ defined in \eqref{eq:Hfreedelay} represent the frequency behavior of the delays operators  $\boldsymbol{\Delta}_{i}$ and $\boldsymbol{\Delta}_{o}$, receptively.
%associate   are defined as  $(\boldsymbol{\Delta}_{i}\Big(\mathbf{u}(t))\Big)_k =  \Big(\mathbf{u}(t-\hat{\tau}_k))_k$, for $k =1,\dots, n_u$ and $\boldsymbol{\Delta}_{i}\Big(\mathbf{C}\mathbf{x}(t))\Big)_j = \Big(\mathbf{C}\mathbf{x}(t-\gamma_j)\Big)_j$, for $j = 1,\dots,n_y$.
The transfer function of the underlying system~\eqref{eq:DynamicalSystemInputDelay} from input $\mathbf{\hat{u}}(t) $ to output $\mathbf{\hat{y}}(t)$ vectors is given by:
\begin{equation}
\hat{\mathbf{H}}_{d}(s)=\hat{\boldsymbol{\Delta}}_{o}(s)\hat{\mathbf{H}}(s)\hat{\boldsymbol{\Delta}}_{i}(s)\in\mathcal{H}_{2}^{\nynu},
\label{eq:TransferFunctionInputDelay}
\end{equation}
where:
\begin{equation}
\left\{\begin{array}{rcl}
\hat{\mathbf{H}}(s)&=&\hat{\mathbf{C}}(\hat{\mathbf{E}}s-\hat{\mathbf{A}})^{-1}\hat{\mathbf{B}}\in{\mathcal{H}_{2}^{\nynu}}\\
%\hat{\mathbf{B}}&=&[\hat{\mathbf{b}}_{1}~\hat{\mathbf{b}}_{2}~\ldots~\hat{\mathbf{b}}_{n_{u}}]\in\mathbb{R}^{n\times{n_{u}}}\\
%\hat{\mathbf{C}}^{T}&=&[\hat{\mathbf{c}}_{1}~\hat{\mathbf{c}}_{2}~\ldots~\hat{\mathbf{c}}_{n_{y}}]\in\mathbb{R}^{n\times{n}_{y}}\\
\hat{\boldsymbol{\Delta}}_{i}(s)&=&\diag(e^{-s\hat{\tau}_{1}}~\ldots~e^{-s\hat{\tau}_{n_{u}}})\in\mathcal{H}_{\infty}^{\nunu}\\
\hat{\boldsymbol{\Delta}}_{o}(s)&=&\diag(e^{-s\hat{\gamma}_{1}}~\ldots~e^{-s\hat{\gamma}_{n_{y}}})\in\mathcal{H}_{\infty}^{\nyny}.
\end{array}\right.
\label{eq:Hfreedelay}
\end{equation}
From this point, we will denote by $\hat{\mathbf{H}}_{d}=(\hat{\mathbf{E}},\hat{\mathbf{A}},\hat{\mathbf{B}},\hat{\mathbf{C}},\hat{\boldsymbol{\Delta}}_{i},\hat{\boldsymbol{\Delta}}_{o})$ a {\bf MIMO} input/output delayed system of the form~\eqref{eq:TransferFunctionInputDelay}. $\hat{\mathbf{H}}_{d}$ will also be said to have order $n\ll N$ (where $N$ is the original model order).

\paragraph{Problem statement} \label{s1ss2}
The main objective addressed in this paper is to solve the following $\mathcal{H}_{2}$ approximation problem:
\begin{prob}{{\bf (Delay model $\mathcal{H}_{2}$-optimal approximation)}}
Given a stable $N^{th}$ order system $\mathbf{G}\in\mathcal{H}_{2}^{\nynu}$, find a reduced $n^{th}$ order (\emph{s.t.} $n\ll N$) multiple-input/output delays model $\hat{\mathbf{H}}_{d}^{\star}=(\hat{\mathbf{E}},\hat{\mathbf{A}},\hat{\mathbf{B}},\hat{\mathbf{C}},\hat{\boldsymbol{\Delta}}_{i},\hat{\boldsymbol{\Delta}}_{o})$ s.t.:
\begin{displaymath}
%\begin{array}{c}
\hat{\mathbf{H}}_{d}^{\star}=\argmin_{\scriptsize\begin{array}{c}\hat{\mathbf{H}}_{d}\in{\mathcal{H}_{2}^{\nynu}}\\
\dim(\hat{\mathbf{H}}_{d})\leq{n}\end{array}} \|\mathbf{G}-\hat{\mathbf{H}}_{d}\|_{\mathcal{H}_{2}},%\\
%~\\
%\text{where:}~\hat{\mathbf{H}}_{d}=\hat{\boldsymbol{\Delta}}_{o}\hat{\mathbf{H}}\hat{\boldsymbol{\Delta}}_{i}~\text{as in~\eqref{eq:TransferFunctionInputDelay}}
%\end{array}
\end{displaymath}
where $\hat{\mathbf{H}}_{d}=\hat{\boldsymbol{\Delta}}_{o}\hat{\mathbf{H}}\hat{\boldsymbol{\Delta}}_{i}$ as in,~\eqref{eq:TransferFunctionInputDelay}.
\label{pb:ProblemMIMO}
\end{prob}
This search for an optimal solution will be carried out assuming that both $\mathbf{G}$ and $\hat{\mathbf{H}}$ from Eq.~\eqref{eq:Hfreedelay} have semi-simple poles {\it i.e.,} \emph{s.t.} their respective transfer function matrix can be decomposed as follows:
\begin{equation}
\mathbf{G}(s)=\sum_{j=1}^{N}\frac{\mathbf{l}_{j}\mathbf{r}_{j}^{T}}{s-\mu_{j}}~\text{and}~\hat{\mathbf{H}}(s)=\sum_{k=1}^{n}\frac{\hat{\mathbf{c}}_{k}\hat{\mathbf{b}}_{k}^{T}}{s-\hat{\lambda}_{k}},
\label{eq:GHmodels}
\end{equation}
where $\forall{j}=1\ldots{N},~\forall{k}=1\ldots{n},~\mathbf{r}_{j},\hat{\mathbf{b}}_{k}\in{\mathbb{C}^{n_{u}\times{1}}}$ and $\mathbf{l}_{j},\hat{\mathbf{c}}_{k}\in{\mathbb{C}^{n_{y}\times{1}}}$. The poles $\mu_{j},\hat{\lambda}_{k}$ are elements of $\mathbb{C_{-}}$ so that $\mathbf{G}$ and $\mathbf{\hat{H}}$ belong to $\mathcal{H}_{2}^{\nynu}$.

%%%%%%%%%%%%%%%%%%%%%%%%%%%%%%%%%%%%%%%%%%%%%%%%%%%%%%%%%%%%%%%%%%%%%%%%%%%%%%%%%
%%%%%%%%%%%%%%%%%%%%%%%%%%%%%%%%%%%%%%%%%%%%%%%%%%%%%%%%%%%%%%%%%%%%%%%%%%%%%%%%%

\section{Preliminary results }\label{sec:Pre}

In this section, some elementary but important, results, which will be useful along this paper, are recalled and generalized.

First of all, a fundamental result dealing with the $\mathcal{H}_{2}$ norm invariance in case of input/output delayed systems is presented.
\begin{prop}{{\bf ($\mathcal{H}_{2}$ norm invariance)}}
Let $\hat{\mathbf{H}}\in\mathcal{H}_{2}^{\nynu}$ be a stable dynamical system and $\mathbf{M}\in\mathcal{H}_{\infty}^{\nunu}$, $\mathbf{N}\in\mathcal{H}_{\infty}^{\nyny}$ s.t.: \begin{equation}\label{eq:CondInvarianceH2}
\forall{\omega}\in\mathbb{R},~\overline{\mathbf{M}(i\omega)}\mathbf{M}(i\omega)^{T}=\mathbf{I}_{n_{u}},~\mathbf{N}(i\omega)^{T}\overline{\mathbf{N}(i\omega)}=\mathbf{I}_{n_{y}}. 
\end{equation}
If $\hat{\mathbf{H}}_{d}=\mathbf{N}\hat{\mathbf{H}}\mathbf{M}$~then~$\|\hat{\mathbf{H}}_{d}\|_{\mathcal{H}_{2}}=\|\hat{\mathbf{H}}\|_{\mathcal{H}_{2}}$.
\label{prop:InvarianceH2}
\end{prop}
\noindent{\it Proof.}~If $\hat{\mathbf{H}}_{d}=\mathbf{N}\hat{\mathbf{H}}\mathbf{M}$, the scaled term $2\pi\|\hat{\mathbf{H}}_{d}\|_{\mathcal{H}_{2}}^{2}$ will then read by definition:
\begin{equation*}
\begin{array}{l}
\displaystyle\int_{-\infty}^{+\infty}\hspace{-0.2in}\trace\left(\overline{\mathbf{N}(i\omega)}\overline{\hat{\mathbf{H}}(i\omega)}\overline{\mathbf{M}(i\omega)}\mathbf{M}^{T}(i\omega)\hat{\mathbf{H}}^{T}(i\omega)\mathbf{N}^{T}(i\omega)\right)\domega\\
=\displaystyle\int_{-\infty}^{+\infty}\hspace{-0.2in}\trace\left(\overline{\mathbf{N}(i\omega)}\overline{\hat{\mathbf{H}}(i\omega)}\hat{\mathbf{H}}^{T}(i\omega)\mathbf{N}^{T}(i\omega)\right)\domega\\
=\displaystyle\int_{-\infty}^{+\infty}\hspace{-0.2in}\trace\left(\overline{\hat{\mathbf{H}}(i\omega)}\hat{\mathbf{H}}^{T}(i\omega)\mathbf{N}^{T}(i\omega)\overline{\mathbf{N}(i\omega)}\right)\domega\\
=\displaystyle\int_{-\infty}^{+\infty}\hspace{-0.2in}\trace\left(\overline{\mathbf{\hat{H}}(i\omega)}\,\mathbf{\hat{H}}(i\omega)^T\right)\domega = 2\pi\| \mathbf{\hat{H}} \|_{\mathcal{H}_2}^{2}.\hfill\square
\end{array}
\end{equation*}
One can easily check that condition~\eqref{eq:CondInvarianceH2} appearing in Proposition~\ref{prop:InvarianceH2} is satisfied by the delays matrices of the two last lines of~\eqref{eq:Hfreedelay} when $\mathbf{M}=\hat{\boldsymbol{\Delta}}_{i}$ and $\mathbf{N}=\hat{\boldsymbol{\Delta}}_{o}$. In other words, the $\mathcal{H}_{2}$ norm does not depend on the input, nor output delays. The following proposition makes now explicit the calculation of the $\mathcal{H}_{2}$ norm associated with the dynamical mismatch gap $\mathbf{G}-\hat{\mathbf{H}}_{d}$, which conditions Problem~\ref{pb:ProblemMIMO} criterion.
\begin{prop} Let $\mathbf{G},~\hat{\mathbf{H}}_{d}\in{\mathcal{H}_{2}^{\nynu}}$ s.t. $\hat{\mathbf{H}}_{d}$ is given by Eq.~\eqref{eq:TransferFunctionInputDelay}. The $\mathcal{H}_{2}$ norm of the approximation gap (or mismatch error), denoted by $\mathcal{J}$, can be expressed as:
\begin{equation}
\begin{array}{rcl}
\mathcal{J}&=&\|\mathbf{G}-\hat{\boldsymbol{\Delta}}_{o}\hat{\mathbf{H}}\hat{\boldsymbol{\Delta}}_{i}\|_{\mathcal{H}_2}^{2}\\
&=&\|\mathbf{G}\|_{\mathcal{H}_{2}}^{2}-2\langle\mathbf{G},\hat{\boldsymbol{\Delta}}_{o}\hat{\mathbf{H}}\hat{\boldsymbol{\Delta}}_{i}\rangle_{\mathcal{H}_{2}}+\|\hat{\mathbf{H}}\|_{\mathcal{H}_{2}}^{2}.
\end{array}
\label{eq:error}
\end{equation}
\label{prop:H2gap} 
\end{prop}
\noindent {\it Proof.}~Simply develop the $\mathcal{H}_{2}$ norm using the inner product definition and exploit the previous result $\|\hat{\boldsymbol{\Delta}}_{o}\hat{\mathbf{H}}\hat{\boldsymbol{\Delta}}_{i}\|_{\mathcal{H}_{2}}=\|\hat{\mathbf{H}}\|_{\mathcal{H}_{2}}$.\hfill$\square$

Obviously, regarding Eq.~\eqref{eq:error}, minimizing $\mathcal{J}$ is equivalent to minimize $-2\langle\mathbf{G},\hat{\boldsymbol{\Delta}}_{o}\hat{\mathbf{H}}\hat{\boldsymbol{\Delta}}_{i}\rangle_{\mathcal{H}_{2}}+\|\hat{\mathbf{H}}\|_{\mathcal{H}_{2}}^{2}$ and thus to look for the optimal values of the decision variables contained in both the realization $\hat{\mathbf{H}}\in\mathcal{H}_{2}^{\nynu}$ and the delay blocks $\hat{\boldsymbol{\Delta}}_{i},~\hat{\boldsymbol{\Delta}}_{o}\in\mathcal{H}_{\infty}^{\nynu}$. At this point, it could be profitable to derive suitable analytical expressions for the inner-product and the $\mathcal{H}_{2}$ norm of $\hat{\mathbf{H}}$ in order to define more precisely the aforementioned $\mathcal{H}_{2}$ gap between the two transfer functions. To this aim, the previous assumption made for both $\mathbf{G}$ and $\hat{\mathbf{H}}$ systems (see Eq.~\eqref{eq:GHmodels}) will be essential to obtain the following results.
\begin{prop}{{\bf ($\mathcal{H}_{2}$ inner product computation with input/output delays)}}
Let $\mathbf{G},~\hat{\mathbf{H}}$ be two systems $\in{\mathcal{H}_{2}^{\nynu}}$ whose respective transfer functions $\mathbf{G}(s)$ and $\hat{\mathbf{H}}(s)$ can be expressed as in~\eqref{eq:GHmodels}. Let $\hat{\boldsymbol{\Delta}}_{i},~\hat{\boldsymbol{\Delta}}_{o}$ be real, $\mathcal{H}_{\infty}^{\nunu}$ and $\mathcal{H}_{\infty}^{\nyny}$ respectively, models satisfying $\sup\{\|\hat{\boldsymbol{\Delta}}_{o}(s),\|\hat{\boldsymbol{\Delta}}_{i}(s)\|/s\in{\mathbb{C}_{-}}\}=M<+\infty$. By denoting $\hat{\mathbf{H}}_{d}=\hat{\boldsymbol{\Delta}}_{o}\hat{\mathbf{H}}\hat{\boldsymbol{\Delta}}_{i}$, the inner product $\langle\hat{\mathbf{H}}_{d},\mathbf{G}\rangle_{\mathcal{H}_{2}}$ reads:
\begin{equation}\label{eq:H2iproductdelay}
\begin{array}{ll}
\langle\hat{\mathbf{H}}_{d},\mathbf{G}\rangle_{\mathcal{H}_{2}}\hspace{-0.1in}&=\mysum_{j=1}^{N}\trace\left(\Res\Big\lbrack\hat{\mathbf{H}}_{d}(-s)\mathbf{G}^{T}(s),\mu_{j}\Big\rbrack\right)\\
&=\mysum_{j=1}^{N}\mathbf{l}_{j}^{T}\hat{\boldsymbol{\Delta}}_{o}(-\mu_{j})\hat{\mathbf{H}}(-\mu_{j})\hat{\boldsymbol{\Delta}}_{i}(-\mu_{j})\mathbf{r}_{j}.
\end{array}
\end{equation} 
\label{prop:InnerProduct}
\end{prop}
\noindent{\it Proof.}~Observing that the poles of the complex function $\hat{\mathbf{H}}_{d}(-s)\mathbf{G}(s)$ are $\mu_{1},~\mu_{2},~\ldots,~\mu_{N}\in{\mathbb{C}_{-}}$ and $-\hat{\lambda}_{1},~{-\hat{\lambda}_{2}},~\ldots,~{-\hat{\lambda}_{n}}\in{\mathbb{C}_{+}}$, let us consider the following semi-circular contour $\Gamma_{C}$ located in the left half plane \emph{s.t.}:
\[\Gamma_{C}=\Gamma_{I}\cup{\Gamma_{R}}\]
with:~$\left\{\begin{array}{l}
\Gamma_{I}=\{s\in{\mathbb{C}}/s=i\omega~\textnormal{and}~\omega\in{[-R;R]},~R\in\mathbb{R}_{+}\}\\
\Gamma_{R}=\{s\in{\mathbb{C}}/s=Re^{i\theta}~\textnormal{where}~\theta\in{\lbrack\pi/2;3\pi/2\rbrack}\}
\end{array}\right..$\\

\noindent Thus, for a sufficient large radius value $R$, the $\Gamma_{C}$ contour will contain all the poles of the transfer function $\mathbf{G}(s)$ {\it i.e.,} $\mu_{1},~\mu_{2},~\ldots,~\mu_{N}$. Thus, by applying the residues theorem, it follows that:
\begin{equation*}
\begin{array}{rcl}
\langle\hat{\mathbf{H}}_{d},\mathbf{G}\rangle_{\mathcal{H}_{2}}&=&\displaystyle\frac{1}{2\pi}\int_{-\infty}^{+\infty}\trace\left(\overline{\hat{\mathbf{H}}_{d}(i\omega)}\mathbf{G}^{T}(i\omega)\right)\domega\\
&=&\displaystyle\lim\limits_{R\rightarrow{+\infty}}\frac{1}{2i\pi}\int_{\Gamma_{C}}\hat{\mathbf{H}}_{d}(-s)\mathbf{G}(s)\ds\\
&=&\mysum_{j=1}^{N}\trace\left(\Res\Big\lbrack\hat{\mathbf{H}}_{d}(-s)\mathbf{G}^{T}(s),\mu_{j}\Big\rbrack\right).
\end{array}
\end{equation*}
where $\Res(.)$ denotes the residue operator. The second equality line holds true since:
\begin{equation*}
\displaystyle\int_{\Gamma_{R}}\big\|\hat{\mathbf{H}}_{d}(-s)\mathbf{G}(s)\ds\big\|\leq{M^{2}\hspace{-0.075in}\int_{\Gamma_{R}}\big\|\hat{\mathbf{H}}(-s)\mathbf{G}(s)ds\big\|}\rightarrow{0}^{+},
\end{equation*}
when $R\rightarrow{+\infty}$.\hfill$\square$\\

One may note that Proposition~\ref{prop:InnerProduct} is a generalization of Lemma 3.5 appearing in~\cite{gugercin2008h_2} in the case of {\bf MIMO} systems with multiple-input/output delays. It is noteworthy that the $\hat{\boldsymbol{\Delta}}_{i}$,  $\hat{\boldsymbol{\Delta}}_{o}$ matrices defined by~\eqref{eq:Hfreedelay} clearly verifies the hypothesis Proposition~\ref{prop:InnerProduct}.
\begin{remark}{{\bf (Delay-free case ''symmetry'')}}
An equivalent proposition was derived in the delay-free case~\cite{gugercin2008h_2}. It can be recovered from Proposition~\ref{prop:InnerProduct} by taking $\hat{\boldsymbol{\Delta}}_{i}=\mathbf{I}_{n_{u}}$ and $\hat{\boldsymbol{\Delta}}_{o}=\mathbf{I}_{n_{y}}$. The result corresponds to the symmetric expression of the inner product {\it i.e.,} the evaluation of $\mathbf{G}$ in the poles of $\hat{\mathbf{H}}$ and its associated residues $\hat{\mathbf{c}}_{k}$ and $\hat{\mathbf{b}}_{k}$ s.t.:
\[\langle\mathbf{G},\hat{\mathbf{H}}\rangle_{\mathcal{H}_{2}}\hspace{-0.025in}=\hspace{-0.05in}\mysum_{k=1}^{n}\hat{\mathbf{c}}_{k}^{T}\hat{\mathbf{G}}(-\hat{\lambda}_{k})\hat{\mathbf{b}}_{k}\hspace{-0.025in}=\hspace{-0.05in}\mysum_{j=1}^{N}\mathbf{l}_{j}^{T}\hat{\mathbf{H}}(-\mu_{j})\mathbf{r}_{j}\hspace{-0.025in}=\hspace{-0.025in}\langle\hat{\mathbf{H}},\mathbf{G}\rangle_{\mathcal{H}_{2}}.\]
In the presence of input/output delays, since the $\mathcal{H}_{2}$ norm cannot be approximated using one contour containing the poles of $\hat{\mathbf{H}}_{d}$ only, this result is no longer true. Indeed, it can be easily shown that in this case, the integral on $\Gamma_{R}$ will depend on a positive exponential argument which will not converge to $0^{+}$ when $R\rightarrow{+\infty}$. This justifies the assumption that $\sup\{\|\hat{\boldsymbol{\Delta}}_{o}(s),\|\hat{\boldsymbol{\Delta}}_{i}(s)\|/s\in{\mathbb{C}_{-}}\}=M<+\infty$ and relevance of Proposition~\ref{prop:InnerProduct}.
\end{remark}
Finally, let us recall the pole(s)/residue(s) $\mathcal{H}_{2}$ norm formula.
\begin{cor}{{\bf (Poles/residues $\mathcal{H}_2$ norm~\cite{gugercin2008h_2})}}
Assume that $\hat{\mathbf{H}}_{d}(s),~\hat{\mathbf{H}}(s)$ belong to $\mathcal{H}_{2}^{\nynu}$ and that $\hat{\mathbf{H}}_{d}=\hat{\boldsymbol{\Delta}}_{o}\hat{\mathbf{H}}\hat{\boldsymbol{\Delta}}_{i}$. Besides, suppose that $\hat{\mathbf{H}}$ can be expressed such as in~\eqref{eq:GHmodels}, then,
\begin{equation*}
\big\|\hat{\mathbf{H}}_{d}\big\|_{\mathcal{H}_{2}}^{2}=\mysum_{k=1}^{n}\hat{\mathbf{c}}_{k}^{T}\hat{\mathbf{H}}(-\hat{\lambda}_{k})\hat{\mathbf{b}}_{k}.
\end{equation*}
\label{cor:H2normPoleResidue}
\end{cor}
\noindent{\it Proof.}~See \cite{gugercin2008h_2}.\hfill$\square$

In the next section, the main result, namely $\mathcal{H}_{2}$ optimality conditions related to Problem~\ref{pb:ProblemMIMO}, are firstly established and an interpolation-based algorithm is proposed to numerically compute the approximation $\hat{\mathbf{H}}_{d}$.

%%%%%%%%%%%%%%%%%%%%%%%%%%%%%%%%%%%%%%%%%%%%%%%%%%%%%%%%%%%%%%%%%%%%%%%%%%%%%%%%%
%%%%%%%%%%%%%%%%%%%%%%%%%%%%%%%%%%%%%%%%%%%%%%%%%%%%%%%%%%%%%%%%%%%%%%%%%%%%%%%%%

\section{Approximation by multiple I/O delays MIMO systems: $\mathcal{H}_{2}$ optimality conditions}\label{sec:MainResult}

Considering the mathematical formulation of Problem~\ref{pb:ProblemMIMO} and the reduced order system structure $\hat{\mathbf{H}}_{d}=\hat{\boldsymbol{\Delta}}_{o}\hat{\mathbf{H}}\hat{\boldsymbol{\Delta}}_{i}$, where $\hat{\mathbf{H}}(s)$ is given as in~\eqref{eq:GHmodels}, the underlying optimization issue that must be solved is parameterized by ($k=1,\dots ,n$): \emph{(i)} the $n$ pole(s) $\hat{\lambda}_{k}\in{\mathbb{C}_{-}}$; \emph{(ii)} the $n$ bi-tangential directions $(\hat{\mathbf{b}}_{k},\hat{\mathbf{c}}_{k})\in\mathbb{C}^{n_{u}\times{1}}\times{\mathbb{C}^{n_{y}\times{1}}}$; and \emph{(iii)} the $n_{u}+n_{y}$ delay values $(\hat{\tau}_{l},\hat{\gamma}_{m}),~l=1\ldots{n_{u}},~m=1\ldots{n_{y}}$. Our primary objective consists in rewriting the expression of the $\mathcal{H}_2$ gap $\mathcal{J}$ as a function of these latter parameters which will subsequently facilitate the derivation of the $\mathcal{H}_{2}$ optimality conditions for Problem~\ref{pb:ProblemMIMO}. This forms the topic of the three following propositions and of Theorem~\ref{theo:MIMOoptimH2}, which stands as the main result of the paper.
\begin{prop} From the preliminary results, the mismatch $\mathcal{H}_{2}$ gap defined previously in Proposition~\ref{prop:H2gap} can be equivalently rewritten as:
\begin{equation} 
\begin{array}{lll}
\mathcal{J}\hspace{-0.1in}&=&\hspace{-0.1in}\|\mathbf{G}\|^2_{\mathcal{H}_2}\hspace{-0.05in}+\mysum_{k=1}^{n}\hat{\mathbf{c}}_{k}^{T}\hat{\mathbf{H}}(-\hat{\lambda}_{k})\hat{\mathbf{b}}_{k}\ldots\\
&&-2\mysum_{j=1}^{N}\mathbf{l}_{j}^{T}\hat{\boldsymbol{\Delta}}_{o}(-\mu_{j})\hat{\mathbf{H}}(-\mu_{j})\hat{\boldsymbol{\Delta}}_{i}(-\mu_{j})\mathbf{r}_{j}.
\end{array}
\label{eqn:J}
\end{equation}
\end{prop}
\noindent {\it Proof.}~The result is immediate. To be established, it requires to develop the $\mathcal{H}_{2}$ norm expression showing the inner product and then to use both Proposition~\ref{prop:InnerProduct} and Corollary~\ref{cor:H2normPoleResidue} results.\hfill$\square$

From the previous equation~\eqref{eqn:J}, the first-order optimality conditions related to the minimization of $\mathcal{J}$ can be analytically computed. The gradient expressions of the $\mathcal{H}_{2}$ gap \emph{w.r.t.} each parameters (delays, tangential directions and poles) are detailed in the two following propositions. Starting with the simplest calculations, we first derive the gradient of $\mathcal{J}$ \emph{w.r.t.} the delays since the second term of the right-hand side part of~\eqref{eqn:J} is delay-dependent, only.
\vspace{-0.1cm}
\begin{prop} The gradients of the $\mathcal{H}_{2}$ gap $\mathcal{J}$ with respect to the delays read $\forall{l=1\ldots{n_{u}}},~\forall{m=1\ldots{n_{y}}}$: 
\begin{equation*}
\begin{array}{c}
\left\{\begin{array}{lll}
\nabla_{\hat{\tau}_{l}}\mathcal{J}&=&-2\displaystyle\frac{\partial{\langle\hat{\mathbf{H}}_{d},\mathbf{G}\rangle_{\mathcal{H}_{2}}}}{\partial{\hat{\tau}_{l}}}\\
&=&-2\mysum_{j=1}^{N}\mu_{j}\mathbf{l}_{j}^{T}\hat{\boldsymbol{\Delta}}_{o}(-\mu_{j})\hat{\mathbf{H}}(-\mu_{j})\mathbf{D}_{l}\hat{\boldsymbol{\Delta}}_{i}(-\mu_{j})\mathbf{r}_{j},\\
\nabla_{\hat{\gamma}_{m}}\mathcal{J}&=&-2\displaystyle\frac{\partial{\langle\hat{\mathbf{H}}_{d},\mathbf{G}\rangle_{\mathcal{H}_{2}}}}{\partial{\hat{\gamma}_{m}}}\\
&=&-2\mysum_{j=1}^{N}\mu_{j}\mathbf{l}_{j}^{T}\mathbf{D}_{m}\hat{\boldsymbol{\Delta}}_{o}(-\mu_{j})\hat{\mathbf{H}}(-\mu_{j})\hat{\boldsymbol{\Delta}}_{i}(-\mu_{j})\mathbf{r}_{j},
\end{array}\right.
\end{array}
\end{equation*}
where elements of $\mathbf{D}_{l}\in\mathbb{R}^{\nunu}$, $\mathbf{D}_{m}\in\mathbb{R}^{\nyny}$, are defined as: $$\left[\mathbf{D}_{k}\right]_{ij}=\delta_{ijk}=\left\{\hspace{-0.075in}\begin{array}{l}1~\text{if}~i=j=k\\0~\text{otherwise}\end{array}\right..$$
\end{prop}
\noindent{\it Proof.}~The proof is straightforward to establish since both $\hat{\boldsymbol{\Delta}}_{i}$ and $\hat{\boldsymbol{\Delta}}_{o}$ terms are diagonal matrices and the exponential derivative function is obvious.\hfill$\square$
\begin{prop} The gradients of the $\mathcal{H}_{2}$ gap $\mathcal{J}$ with respect to parameters $\hat{\mathbf{c}}_{k}$, $\hat{\mathbf{b}}_{k}$ and $\hat{\lambda}_{k},~\forall{k=1\ldots{n}}$ read:
\begin{equation*}
\left\{\begin{array}{lll}
\nabla_{\hat{\mathbf{c}}_{k}}\mathcal{J}&=&-2\displaystyle\frac{\partial{\langle\hat{\mathbf{H}}_{d},\mathbf{G}\rangle_{\mathcal{H}_{2}}}}{\partial\hat{\mathbf{c}}_{k}}+\displaystyle\frac{\partial{\|\hat{\mathbf{H}}\|_{\mathcal{H}_{2}}^{2}}}{\partial\hat{\mathbf{c}}_{k}}\\
&=&-2\hat{\mathbf{b}}_{k}^T\left(\tilde{\mathbf{G}}(-\hat{\lambda}_{k})-\hat{\mathbf{H}}(-\hat{\lambda}_{k})\right)^T,\\
\nabla_{\hat{\mathbf{b}}_{k}}\mathcal{J}&=&-2\hat{\mathbf{c}}_{k}^T\left(\tilde{\mathbf{G}}(-\hat{\lambda}_{k})-\hat{\mathbf{H}}(-\hat{\lambda}_{k})\right),\\
\nabla_{\hat{\lambda}_{k}}\mathcal{J}&=&2\hat{\mathbf{c}}_{k}^{T}\left(\tilde{\mathbf{G}}^{\prime}(-\hat{\lambda}_{k})-\hat{\mathbf{H}}^{\prime}(-\hat{\lambda}_{k})\right)\hat{\mathbf{b}}_{k},
\end{array}\right.
\end{equation*}
where:
\begin{equation}
\tilde{\mathbf{G}}(s)=\mysum_{j=1}^{N}\hat{\boldsymbol{\Delta}}_{o}(-\mu_{j})\displaystyle\frac{\mathbf{l}_{j}^{T}\mathbf{r}_{j}}{s-\mu_{j}}\hat{\boldsymbol{\Delta}}_{i}(-\mu_{j}).
\label{eq:Gtilde}
\end{equation}
and where $\tilde{\mathbf{G}}^{\prime}$ and ${\mathbf{\hat{H}}}^{\prime}$ are the Laplace derivative of $\tilde{\mathbf{G}}$ and ${\mathbf{\hat{H}}}$, respectively.
\end{prop}
\noindent{\it Proof.}~By defining $\tilde{\mathbf{r}}_{j}=\hat{\boldsymbol{\Delta}}_{i}(-\mu_{j})\mathbf{r}_{j}$ and $\tilde{\mathbf{l}}_{j}^{T}=\mathbf{l}_{j}^{T}\hat{\boldsymbol{\Delta}}_{o}(-\mu_{j})$ with $j=1\ldots{N}$, the $\mathcal{H}_{2}$ gap can be written as: 
\begin{displaymath}
\begin{array}{lll}
\mathcal{J}&=&\|\mathbf{G}\|_{\mathcal{H}_{2}}^{2}-2\mysum_{j=1}^{N}\tilde{\mathbf{l}}_{j}^{T}\Big(\displaystyle\sum_{m=1}^{n}\displaystyle\frac{\hat{\mathbf{c}}_{m}\hat{\mathbf{b}}_{m}^{T}}{-\mu_j-\hat{\lambda}_{m}}\Big)\tilde{\mathbf{r}}_{j}\\ &+&\mysum_{k=1}^{n}\hat{\mathbf{c}}_{k}^{T}\Big(\displaystyle \sum_{m=1}^{n}\displaystyle\frac{\hat{\mathbf{c}}_{m}\hat{\mathbf{b}}_{m}^{T}}{-\hat{\lambda}_{k}-\hat{\lambda}_{m}}\Big)\hat{\mathbf{b}}_{k}.
\end{array}
\end{displaymath}

Then, calculating the gradients \emph{w.r.t.} $\hat{\mathbf{b}}_{l},~\hat{\mathbf{c}}_{l}$ and $\hat{\lambda}_{l}~(l=1\ldots{n})$ gives:
\begin{equation*}
\nabla_{\hat{\mathbf{b}}_{l}}\mathcal{J} = \displaystyle -2\frac{\partial{\langle\hat{\mathbf{H}}_{d},\mathbf{G}\rangle_{\mathcal{H}_{2}}}}{\partial\hat{\mathbf{b}}_{l}}+\displaystyle\frac{\partial{\|\hat{\mathbf{H}}\|_{\mathcal{H}_{2}}^{2}}}{\partial\hat{\mathbf{b}}_{l}}
\end{equation*}
Thus, by computing both terms on this expression
\begin{equation*}
\begin{array}{lll}
\displaystyle\frac{\partial{\|\hat{\mathbf{H}}\|_{\mathcal{H}_{2}}^{2}}}{\partial\hat{\mathbf{b}}_{l}} &=& \mysum_{k=1}^{n} \mysum_{m=1}^{n}   \displaystyle\frac{(\hat{\mathbf{c}}_{k}^{T}\hat{\mathbf{c}}_{m}) }{-\hat{\lambda}_{k}-\hat{\lambda}_{m}} \frac{\partial{~}}{\partial\hat{\mathbf{b}}_{l}}\Big(\hat{\mathbf{b}}_{m}^{T}\hat{\mathbf{b}}_{k}\Big) \\
&=& 2 \mysum_{k=1}^{n}   \displaystyle\frac{\hat{\mathbf{c}}_{l}^{T}\hat{\mathbf{c}}_{k}\hat{\mathbf{b}}_{k}^T  }{-\hat{\lambda}_{k}-\hat{\lambda}_{l}}  = 2\hat{\mathbf{c}}_{l}^{T} \hat{\mathbf{H}}(-\hat{\lambda}_{l})
\end{array} 
\end{equation*}
and 
\begin{equation*}
\begin{array}{lll}
\displaystyle\frac{\partial{\langle\hat{\mathbf{H}}_{d},\mathbf{G}\rangle_{\mathcal{H}_{2}}}}{\partial\hat{\mathbf{b}}_{l}} &=&  \mysum_{j=1}^{N}\mysum_{m=1}^{n}  \displaystyle\frac{(\mathbf{\tilde{l}}_{j}^{T}\hat{\mathbf{c}}_{m})\mathbf{\tilde{r}}_{j}^T  }{-\mu_j-\hat{\lambda}_{m}}  \nabla_{\hat{\mathbf{b}}_{l}}\hat{\mathbf{b}}_{m} \\
&=&   \hat{\mathbf{c}}_{l}^T\mysum_{j=1}^{N} \displaystyle\frac{\mathbf{\tilde{l}}_{j}\mathbf{\tilde{r}}_{j}^T  }{-\mu_j-\hat{\lambda}_{l}}  =  \hat{\mathbf{c}}_{l}^T \tilde{\mathbf{G}}(-\hat{\lambda}_{l}).
\end{array} 
\end{equation*}
one obtains the gradient.

It is noteworthy that $\nabla_{\hat{\mathbf{c}}_{l}}\mathcal{J}$ can be obtained in the same way as $\nabla_{\hat{\mathbf{b}}_{l}}\mathcal{J}$. The calculation of $\nabla_{\hat{\lambda}_{l}}\mathcal{J}$ is straightforwardly derived as follows:
\begin{equation*}
\begin{array}{lll}
\nabla_{\hat{\lambda}_{l}}\mathcal{J}&=&-2\mysum_{j=1}^{N}\displaystyle\frac{\tilde{\mathbf{l}}_{j}^{T}\hat{\mathbf{c}}_{l}\hat{\mathbf{b}}_{l}^{T}\tilde{\mathbf{r}}_{j}}{(-\hat{\lambda}_{l}-\mu_{j})^2}-\hat{\mathbf{c}}_{l}^{T}\hat{\mathbf{H}}^{\prime}(-\hat{\lambda}_{l})\hat{\mathbf{b}}_{l}\ldots\\
&&+\mysum_{k=1}^{n}\displaystyle\frac{\hat{\mathbf{c}}_{k}^{T}\hat{\mathbf{c}}_{l}\hat{\mathbf{b}}_{l}^{T}\hat{\mathbf{b}}_{k}}{(-\hat{\lambda}_{l}-\hat{\lambda}_{k})^2}\\
&=&2\hat{\mathbf{c}}_{l}^{T}\left(\tilde{\mathbf{G}}^{\prime}(-\hat{\lambda}_{l})-\hat{\mathbf{H}}^{\prime}(-\hat{\lambda}_{l})\right)\hat{\mathbf{b}}_{l}.\hfill\hspace{0.55in}\square
\end{array}
\end{equation*}

Theorem~\ref{theo:MIMOoptimH2} gathers all the first-order optimality conditions related to Problem~\ref{pb:ProblemMIMO} and stands as the main result of the paper.
\begin{theorem}{{\bf (Delay model approximation first-order $\mathcal{H}_2$ optimality conditions)}}
Let us consider $\mathbf{G}\in\mathcal{H}_{2}^{\nynu}$ whose transfer function is $\mathbf{G}(s)\in\mathbb{C}^{\nynu}$. Let $\hat{\mathbf{H}}_{d}=\hat{\boldsymbol{\Delta}}_{o}\hat{\mathbf{H}}\hat{\boldsymbol{\Delta}}_{i}$ be a local optimum of Problem~\ref{pb:ProblemMIMO}. It is assumed that $\hat{\mathbf{H}}\in\mathcal{H}_{2}^{\nynu}$ corresponds to a model with semi-simple poles only and whose transfer function is denoted by $\hat{\mathbf{H}}(s)=\hat{\mathbf{C}}(s\hat{\mathbf{E}}-\hat{\mathbf{A}})^{-1}\hat{\mathbf{B}}\in\mathbb{C}^{\nynu}$. Let $\hat{\boldsymbol{\Delta}}_{i},~\hat{\boldsymbol{\Delta}}_{o}$ be elements of $\mathcal{H}_{\infty}^{\nunu}$ and $\mathcal{H}_{\infty}^{\nyny}$, respectively, s.t. Propositions~\ref{prop:InvarianceH2} and~\ref{prop:InnerProduct} are verified. Then, the following equalities hold:
\begin{equation}
\left\{\begin{array}{l}
\hat{\mathbf{H}}(-\hat{\lambda}_{k})\hat{\mathbf{b}}_{k}=\tilde{\mathbf{G}}(-\hat{\lambda}_{k})\hat{\mathbf{b}}_{k},\\
\hat{\mathbf{c}}_{k}^{T}\hat{\mathbf{H}}(-\hat{\lambda}_{k})=\hat{\mathbf{c}}_{k}^{T}\tilde{\mathbf{G}}(-\hat{\lambda}_{k}),\\
\hat{\mathbf{c}}_{k}^{T}\hat{\mathbf{H}}^{\prime}(-\hat{\lambda}_{k})\hat{\mathbf{b}}_{k}=\hat{\mathbf{c}}_{k}^{T}\tilde{\mathbf{G}}^{\prime}(-\hat{\lambda}_{k})\hat{\mathbf{b}}_{k}, 
\end{array}\right.
\label{eq:H2interpol}
\end{equation}
\begin{equation}
\left\{\begin{array}{l}
\mysum_{j=1}^{N}\mu_{j}\mathbf{l}_{j}^{T}\hat{\boldsymbol{\Delta}}_{o}(-\mu_{j})\hat{\mathbf{H}}(-\mu_{j})\mathbf{D}_{l}\hat{\boldsymbol{\Delta}}_{i}(-\mu_{j})\mathbf{r}_{j}=0,\\
\mysum_{j=1}^{N}\mu_{j}\mathbf{l}_{k}^{T}\mathbf{D}_{m}\hat{\boldsymbol{\Delta}}_{o}(-\mu_{j})\hat{\mathbf{H}}(-\mu_{j})\hat{\boldsymbol{\Delta}}_{i}(-\mu_{j})\mathbf{r}_{j}=0,
\end{array}\right.
\label{eq:H2tau}
\end{equation}
for all $k=1\ldots{n}, l=1\ldots{n_{u}}$ and $m=1\ldots{n_{y}}$ where  $\tilde{\mathbf{G}}(s)$ is given by~\eqref{eq:Gtilde}.
\label{theo:MIMOoptimH2}
\end{theorem}
\noindent{\it Proof.}~The interpolation conditions gathered in~\eqref{eq:H2interpol} are deduced by taking $\nabla_{\hat{\mathbf{c}}_{l}}\mathcal{J}=0$, $\nabla_{\hat{\mathbf{b}}_{l}}\mathcal{J}=0$ and $\nabla_{\hat{\lambda}_{l}}\mathcal{J}=0$. Conditions~\eqref{eq:H2tau} are obtained similarly by taking $\nabla_{\hat{\tau}_{l}}\mathcal{J}=0$ and $\nabla_{\hat{\gamma}_{m}}\mathcal{J}=0$.\hfill$\square$

Theorem~\ref{theo:MIMOoptimH2} asserts that any solution of the $\mathcal{H}_{2}$ model approximation Problem~\ref{pb:ProblemMIMO}, denoted by $\hat{\mathbf{H}}_{d}=\hat{\boldsymbol{\Delta}}_{o}\hat{\mathbf{H}}\hat{\boldsymbol{\Delta}}_{i}$ is \emph{s.t.} $\hat{\mathbf{H}}$ satisfies, at the same time, a set of $3n$ bi-tangential interpolation conditions detailed in~\eqref{eq:H2interpol} and another set of $n_{u}+n_{y}$ relations on the delays contained in the $\hat{\boldsymbol{\Delta}}_{i}$ and $\hat{\boldsymbol{\Delta}}_{o}$ diagonal matrices~\eqref{eq:H2tau}.

\begin{remark}{{\bf ($\mathcal{H}_2$ optimality conditions in the SISO case)}} In the SISO case, all the conditions provided in Theorem~\ref{theo:MIMOoptimH2} appear much simpler and can be stated as follows. Considering:
\begin{displaymath}
\mathbf{G}(s)=\mysum_{j=1}^{N}\displaystyle\frac{\psi_{j}}{s-\mu_{j}},~\hat{\mathbf{H}}_{d}(s)=\mysum_{k=1}^{n}\displaystyle\frac{\phi_{k}e^{-\tau s}}{s-\hat{\lambda}_{k}},
\end{displaymath}
s.t. $\hat{\mathbf{H}}_{d}$ is a local optimum of Problem~\ref{pb:ProblemMIMO}, then the following conditions hold:
\begin{equation}\label{eq:H2interpolSISO}
\left\{\begin{array}{l}
\hat{\mathbf{H}}(-\hat{\lambda}_{k})=\tilde{\mathbf{G}}(-\hat{\lambda}_{k}),\\ 
\hat{\mathbf{H}}^{\prime}(-\hat{\lambda}_{k})=\tilde{\mathbf{G}}^{\prime}(-\hat{\lambda}_{k}),  
\end{array}\right.
\end{equation}
\begin{equation}\label{eq:H2tauSISO}
\begin{array}{l}
\mysum_{j=1}^{N}\mu_{j}\psi_{j}\left(\mysum_{k=1}^{n}\displaystyle\frac{\phi_{k}}{\mu_{j}+\hat{\lambda}_{k}}\right)e^{\tau\mu_{j}} = 0.
\end{array}
\end{equation}
for all $k=1\ldots{n}$, and where $\tilde{\mathbf{G}}$ is as in~\eqref{eq:Gtilde}:
\begin{equation}
\tilde{\mathbf{G}}(s)=\mysum_{j=1}^{N}\displaystyle\frac{\psi_{j}}{s-\mu_{j}}e^{\tau \mu_{j}}.
\nonumber
\end{equation}
\end{remark}

\begin{remark}[Impulse response of $\tilde{\mathbf{G}}(s)$ and advance effect]
The $\Htwo$-optimality conditions given in Theorem \ref{theo:MIMOoptimH2} involves a model $\tilde{\mathbf{G}}(s)$ which has a pole-residue decomposition defined by \eqref{eq:Gtilde}. For simplicity, let us consider the SISO case where  $\mathbf{G}$ and  $\mathbf{\tilde{G}}$ is given by 
\[ \mathbf{G}(s)=\mysum_{j=1}^{N}\displaystyle\frac{\psi_{j}}{s-\mu_{j}}~\,~ \tilde{\mathbf{G}}(s)=\mysum_{j=1}^{N}\displaystyle\frac{\psi_{j}}{s-\mu_{j}}e^{\mu_{j} \tau}.\]
Thus, the the impulse response of $\tilde{\mathbf{G}}(s)$ is 
\begin{equation}
\begin{array}{lll}
\mathbf{\tilde{g}}(t) = \mysum_{j=1}^{N}\displaystyle \psi_j e^{\mu_{j} t}e^{\mu_{j} \tau}\mathbf{1}(t) &=& \mysum_{j=1}^{N}\displaystyle \psi_j e^{\mu_{j}(t+\tau)}\mathbf{1}(t)  \\  &=& \mathbf{g}(t+\tau)\mathbf{1}(t),~\,t \in \mathbb{R} 
\nonumber
\end{array}
\end{equation}
where $\mathbf{1}(t)$ corresponds to the Heaviside step function and $\mathbf{g}(t)$ is the impulse response of model $\mathbf{G}(s)$. Therefore, $\tilde{\mathbf{G}}(s)$ behaves as a time advance of $\mathbf{G}(s)$ and correspond to the "causal part" of the model $\mathbf{G}(s)e^{s\tau}$. 
\end{remark}

\subsection{Practical considerations}\label{subsec:considerantions}
In this subsection, three considerations about Problem~\ref{pb:ProblemMIMO} and Theorem \ref{theo:MIMOoptimH2} are discussed. These latter are relevant to sketch an algorithm which enables the computation of model $\hat{\boldsymbol{\Delta}}_{o}\hat{\mathbf{H}}\hat{\boldsymbol{\Delta}}_{i}$ satisfying the optimality conditions of Theorem~\ref{theo:MIMOoptimH2}. Let us consider that  $\hat{\mathbf{H}}_{d}=\hat{\boldsymbol{\Delta}}_{o}\hat{\mathbf{H}}\hat{\boldsymbol{\Delta}}_{i}$ is a local minimum of the $\mathcal{H}_{2}$ optimization Problem~\ref{pb:ProblemMIMO} where $\hat{\mathbf{H}}$ is given by~\eqref{eq:GHmodels}, then:
\begin{itemize}
%[\ding{224}]
\item \textbf{Consideration \ding{202}.}~If the matrices $\hat{\boldsymbol{\Delta}}_{o},~\hat{\boldsymbol{\Delta}}_{i}$ and the reduced order model poles $\hat{\lambda}_{1},~\hat{\lambda}_{2},~\ldots,~\hat{\lambda}_{n}$ are assumed to be known, Problem~\ref{pb:ProblemMIMO} is reduced to a much simpler problem that can be solved, for example, by using the well-known Loewner framework such as in~\cite{mayo2007framework};
%[\ding{224}]
\item \textbf{Consideration \ding{203}.}~If the delay matrices $\hat{\boldsymbol{\Delta}}_{o},~\hat{\boldsymbol{\Delta}}_{i}$ are known, then Problem~\ref{pb:ProblemMIMO} can be solved by finding a model realization $\hat{\mathbf{H}}$ which satisfies the interpolation conditions~\eqref{eq:H2interpol} of Theorem~\ref{theo:MIMOoptimH2}, only. This can be done using, for instance, a very efficient iterative algorithm, \emph{e.g.,} \textbf{IRKA} (see \cite{gugercin2008h_2});

%[\ding{224}]
\item \textbf{Consideration \ding{204}.}~Assume that the system realization $\hat{\mathbf{H}}$ has already been determined. It follows that Problem~\ref{pb:ProblemMIMO} is equivalent to look for optimal delays matrices $(\hat{\boldsymbol{\Delta}}_{o}^{\star},\hat{\boldsymbol{\Delta}}_{i}^{\star})\in\mathcal{H}_{\infty}^{\nyny}\times\mathcal{H}_{\infty}^{\nunu}$ \emph{s.t.}:
\begin{equation}
(\hat{\boldsymbol{\Delta}}_{o}^{\star},\hat{\boldsymbol{\Delta}}_{i}^{\star})=\argmax_{(\hat{\boldsymbol{\Delta}}_{o},\hat{\boldsymbol{\Delta}}_{i})}\langle\hat{\boldsymbol{\Delta}}_{o}\hat{\mathbf{H}}\hat{\mathbf{\Delta}}_{i},\mathbf{G}\rangle_{\mathcal{H}_{2}}.
\label{eq:optDelta}
\end{equation} 
Interestingly, since $\langle\hat{\boldsymbol{\Delta}}_{o}\hat{\mathbf{H}}\hat{\mathbf{\Delta}}_{i},\mathbf{G}\rangle_{\mathcal{H}_{2}}\rightarrow{0}$ when the delays go to infinity, this problem can be restricted to a compact set and thus a global solution exists.
\end{itemize}  

\subsection{Computational considerations}
An algorithm which allows to numerically compute a model $\hat{\mathbf{H}}_{d}$ satisfying the previous $\mathcal{H}_{2}$ optimality conditions is proposed in this subsection. It relies on the considerations above discussed (Section~\ref{subsec:considerantions}). Therefore, the proposed approach corresponds to an iterative algorithm in which each iteration can be decomposed in two steps. The first one aims at computing a realization $\hat{\mathbf{H}}$ which satisfies the interpolation conditions~\eqref{eq:H2interpol} while fixing the matrices $\hat{\boldsymbol{\Delta}}_{o},~\hat{\boldsymbol{\Delta}}_{i}$ at their values obtained from the previous iteration. This can be done using, for instance, the \textbf{IRKA} algorithm (\emph{Step 4}). In the second step, the resulting $\hat{\mathbf{H}}$ is then exploited to determine the $n_{u}+n_{y}$ optimal values for the $\hat{\boldsymbol{\Delta}}_{o},~\hat{\boldsymbol{\Delta}}_{i}$ matrices elements (\emph{Step 5}). This step is achieved by solving the nonlinear optimization problem defined in~\eqref{eq:optDelta} using an appropriate solver. Then, the whole process is repeated and these two steps performed again until the convergence\footnote{In practice, different stopping criteria might be considered, \emph{e.g.} \emph{(i)} the variation of the interpolation points materialized by $\hat{\lambda}_k$ ($k=1,\dots ,n$), as in \cite{gugercin2008h_2}, \emph{(ii)} the interpolation conditions check (Theorem \ref{theo:MIMOoptimH2}) or \emph{(iii)} the mismatch $\mathcal H_2$ error check (if the order $N$ of the original system is reasonably low).}. At the end of the procedure, the model built will satisfy the $\mathcal{H}_{2}$ optimality conditions on which Theorem~\ref{theo:MIMOoptimH2} relies. This sequential procedure can be summarized such as in Algorithm~\ref{algo:2stages}, and referred to as \textbf{MIMO IO-dIRKA}.  
\begin{algorithm}
\caption{\textbf{MIMO IO-dIRKA} (MIMO Input Output delay IRKA)}
\begin{algorithmic}[1]
\REQUIRE A $N^{th}$-order model $\mathbf{G}\in\mathcal{H}_{2}^{\nynu}$, dimension $n \in\mathbb{N}^{\ast}$ ($n\ll N$) and initial guesses for both $\hat{\boldsymbol{\Delta}}_{i}^{\iter=0},\hat{\boldsymbol{\Delta}}_{o}^{\iter=0}$.
\WHILE{not converged}
\STATE Set $\iter\leftarrow{\iter+1}$
\STATE Build $\tilde{\mathbf{G}}^{\iter}$ as in \eqref{eq:Gtilde}%, with $\hat{\boldsymbol{\Delta}}_{i}^{\iter}=\hat{\boldsymbol{\Delta}}_{i}^{\iter-1},~\hat{\boldsymbol{\Delta}}_{o}^{\iter}=\hat{\boldsymbol{\Delta}}_{o}^{\iter-1}$
\STATE Build $\hat{\mathbf{H}}^{\iter}$ satisfying the bi-tangential interpolation conditions \eqref{eq:H2interpol} using \textbf{IRKA}~\cite{gugercin2008h_2} on $\tilde{\mathbf{G}}^{\iter}$
\STATE Determine $(\hat{\boldsymbol{\Delta}}_{i}^{\star},~\hat{\boldsymbol{\Delta}}_{o}^{\star})$ which solve \eqref{eq:optDelta} using $\hat{\mathbf{H}}^{\iter}$
\STATE Set $\hat{\boldsymbol{\Delta}}_{i}^{\iter}\leftarrow\hat{\boldsymbol{\Delta}}_{i}^{\star}$,~$\hat{\boldsymbol{\Delta}}_{o}^{\iter}\leftarrow\hat{\boldsymbol{\Delta}}_{o}^{\star}$
\ENDWHILE
\STATE Construct  $\hat{\mathbf{H}}_{d}=\hat{\boldsymbol{\Delta}}_{o}^{\iter}\hat{\mathbf{H}}^{\iter}\hat{\boldsymbol{\Delta}}_{i}^{\iter}$
\ENSURE $\hat{\mathbf{H}}_{d}$ satisfies the interpolation conditions of Theorem \ref{theo:MIMOoptimH2}.
\end{algorithmic}
\label{algo:2stages}
\end{algorithm}

\subsection{Structured input/output delays}
All the previous results are left unchanged in the case of structured input/output delays \emph{i.e.,} if, for example, delays does not apply on given input(s) and/or output(s) of $\hat{\mathbf{H}}_{d}$. The results can be derived in a straightforward way, without any loss of generality, just by considering the following ordered delays matrices (where delays are present on the first $n_{d1}<n_u$ inputs and $n_{d2}<n_y$ outputs):
\begin{equation*}
\left\{\begin{array}{l}
\hat{\boldsymbol{\Delta}}_{i}(s)=\diag(e^{-s\hat{\tau}_{1}},~e^{-s\hat{\tau}_{2}},~\ldots,~e^{-s\hat{\tau}_{n_{d1}}},~1,~\ldots,~1)\\
\hat{\boldsymbol{\Delta}}_{o}(s)=\diag(e^{-s\hat{\gamma}_{1}},~e^{-s\hat{\gamma}_{2}},~\ldots,~e^{-s\hat{\gamma}_{n_{d2}}},~1,~\ldots,~1).
\end{array}\right.
\end{equation*}
One can easily note that the preliminary results from Sections~\ref{sec:Pre} and~\ref{sec:MainResult} still remain true when introducing these matrices. The main result stated in Theorem \ref{theo:MIMOoptimH2} thus remains unchanged. %Moreover, the same methodology can be used to solve Problem~\ref{pb:ProblemMIMO}.

\section{Numerical application}\label{sec:applications}

This section is dedicated to the application of the results obtained in Sections \ref{sec:MainResult}, namely, the input/output-delay optimal $\mathcal H_2$ model approximation and its first -order optimality conditions. We will emphasize the potential benefit and effectiveness of the proposed approach.

\begin{figure}[h]
  	\centering
	\includegraphics[width = 8cm]{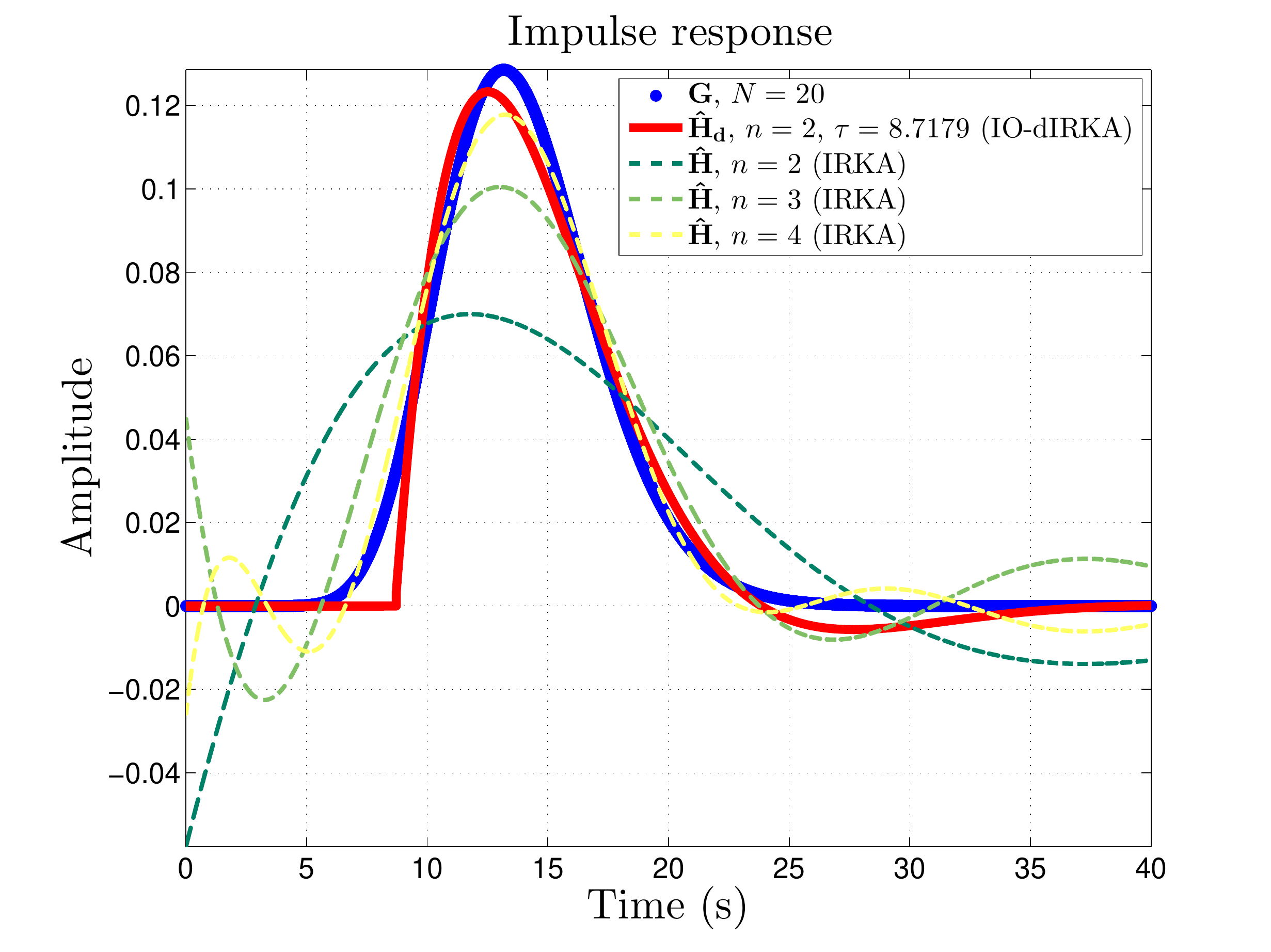}
	 \caption{Impulse response of the original model $\mathbf{H}$ of order $N=20$ (solid dotted blue line), the input-delay $\Htwo$-optimal model $\hat{\mathbf{H}}_{d}$ of order $n=2$ (solid red line) and the delay-free $\Htwo$-optimal models $\hat{\mathbf{H}}$ of order $n= \{2,3,4\}$  (dashed dark green, light green and yellow lines).}
	 \label{fig:Exemple2}
\end{figure}

Let us consider a model $\mathbf{G}$ of order $N=20$, given by the following transfer function
\begin{equation}\label{eq:ModelEx2}
\mathbf{G}(s) = \prod_{j=1}^N \dfrac{\mu_j}{s-\mu_j},
\end{equation}
where $\mu_j\in\mathbb R_-$ ($j=1,\dots,N$) are linearly spaced between $[-2~-1]$. The impulse response of $\mathbf{G}$ is given by the solid dotted blue line in Figure \ref{fig:Exemple2}. Interestingly, it behaves like a system with an input delay. In order to fit the framework proposed in this paper, input-delay $\Htwo$ optimal model $\hat{\mathbf{H}}_{d}=\hat{\boldsymbol{\Delta}}_{o}\hat{\mathbf{H}}\hat{\boldsymbol{\Delta}}_{i}$ of order $n=2$ (solid red) was obtained by applying Theorem \ref{theo:MIMOoptimH2} and \textbf{IO-dIRKA}, as described in Section \ref{sec:MainResult}. The obtained delay model is compared with delay-free approximations of order $n=\{2,3,4\}$, obtained with \textbf{IRKA}\footnote{Using the implementation available in the \textbf{MORE toolbox} \cite{PoussotMORE:2012}, \texttt{http://w3.onera.fr/more/}.}. All the results are reported on  Figure \ref{fig:Exemple2}. 

As clearly shown on Figure \ref{fig:Exemple2}, the proposed methodology allows to obtain an input-delay $\Htwo$ approximation of model $\mathbf{G}$ that clearly provides a better matching than the delay-free cases, even for higher orders (here, \textbf{IRKA} with $n=4$ still have a bad matching and exhibits difficulties in accurately catching the delay and main dynamics). Indeed, the delay-free cases exhibits an oscillatory behaviour during the first seconds while the input-delay model $\hat{\mathbf{H}}_{d}$ takes benefit of the delay structure to focus on the main dynamical effect. Moreover, the approximation model of $\hat{\mathbf{H}}_{d}$ satisfies the conditions given in Theorem \ref{theo:MIMOoptimH2}.

\begin{remark}[Numerical results (SISO case, $n=2$)]
For sake of completeness, the optimal numerical values obtained with \textbf{MIMO IO-dIRKA} are: $\hat{\lambda}_{1,2}= -2.0320\times 10^{-1} \pm i~2.0700\times10^{-1}$, $\hat{\phi}_{1,2}=1.5713\times10^{-3} \pm i~1.8691\times10^{-1}$ and the optimal delay $\tau=8.7179$. The interpolation conditions can then easily be checked: 
\begin{itemize}
\item Condition \eqref{eq:H2interpolSISO} leads to $\hat{\mathbf{H}}(-\hat{\lambda}_{1,2})=\tilde{\mathbf{G}}(-\hat{\lambda}_{1,2})=2.3567\times10^{-1} - i~ 2.3614\times10^{-1}$  and 
$\hat{\mathbf{H}}^{\prime}(-\hat{\lambda}_{1,2})=\tilde{\mathbf{G}}^{\prime}(-\hat{\lambda}_{1,2})= 5.6466\times10^{-1} \pm i~1.1465$.
\item When evaluating $\mysum_{j=1}^{N}\mu_{j}\psi_{j}\left(\mysum_{k=1}^{n}\displaystyle\frac{\phi_{k}}{\mu_{j}+\hat{\lambda}_{k}}\right)e^{\tau\mu_{j}}$, one obtains $9.7284\times 10^{-5}$, which is close to zero, as stated by condition \eqref{eq:H2tauSISO}.
\end{itemize}
\end{remark}

\begin{figure}[H]
  	\centering
	\includegraphics[width = 8cm]{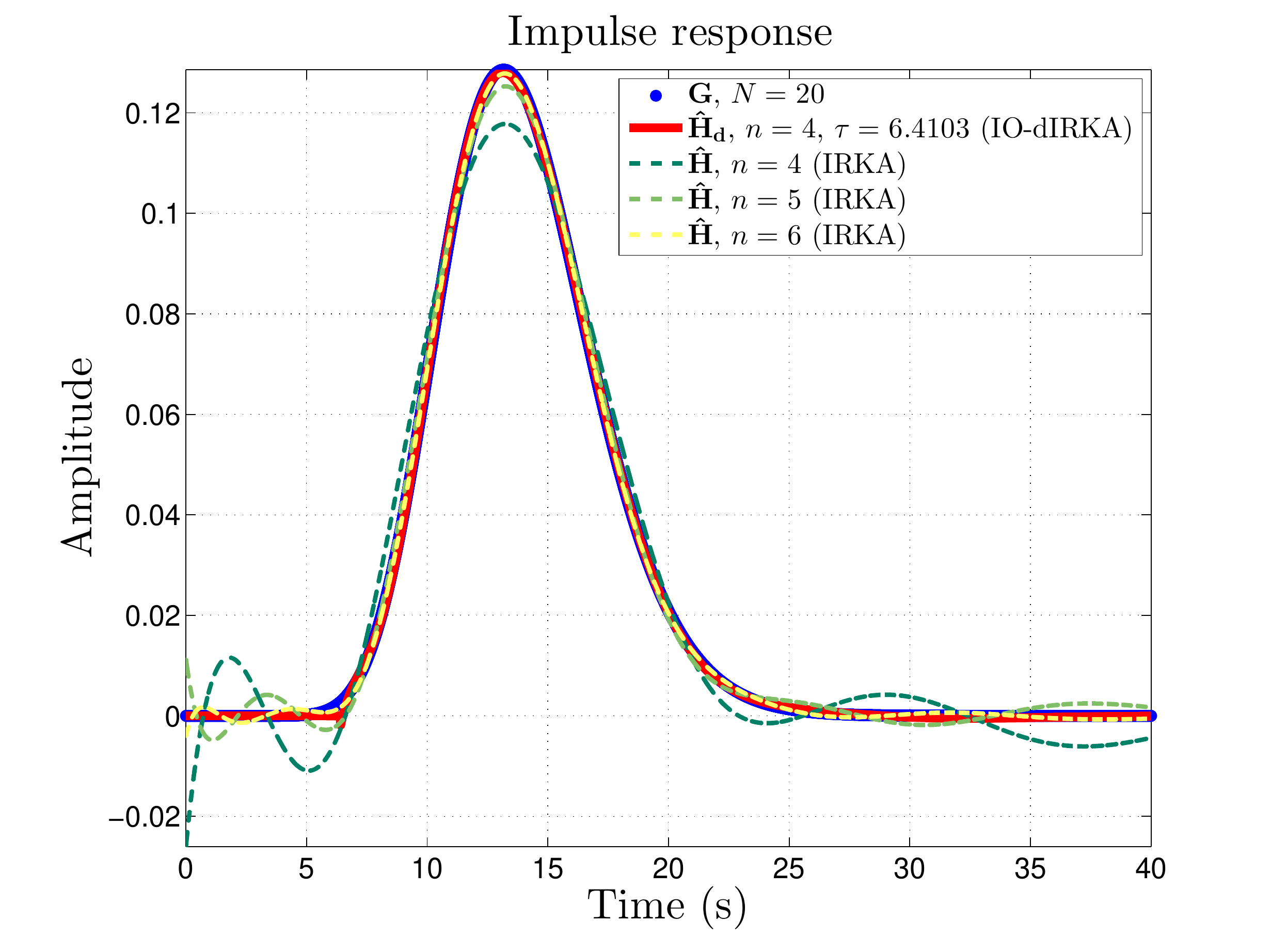}
	 \caption{Impulse response of the original model $\mathbf{H}$ of order $N=20$  (solid dotted blue line), the input-delay $\Htwo$-optimal model $\hat{\mathbf{H}}_{d}$ of order $n=4$ (solid red line) and the delay-free $\Htwo$-optimal models $\hat{\mathbf{H}}$ of order $n= \{4,5,6\}$ (dashed dark green, light green and yellow lines).}
	 \label{fig:Exemple2r4}
\end{figure}

\begin{figure}[H]
  	\centering
	\includegraphics[width = 8cm]{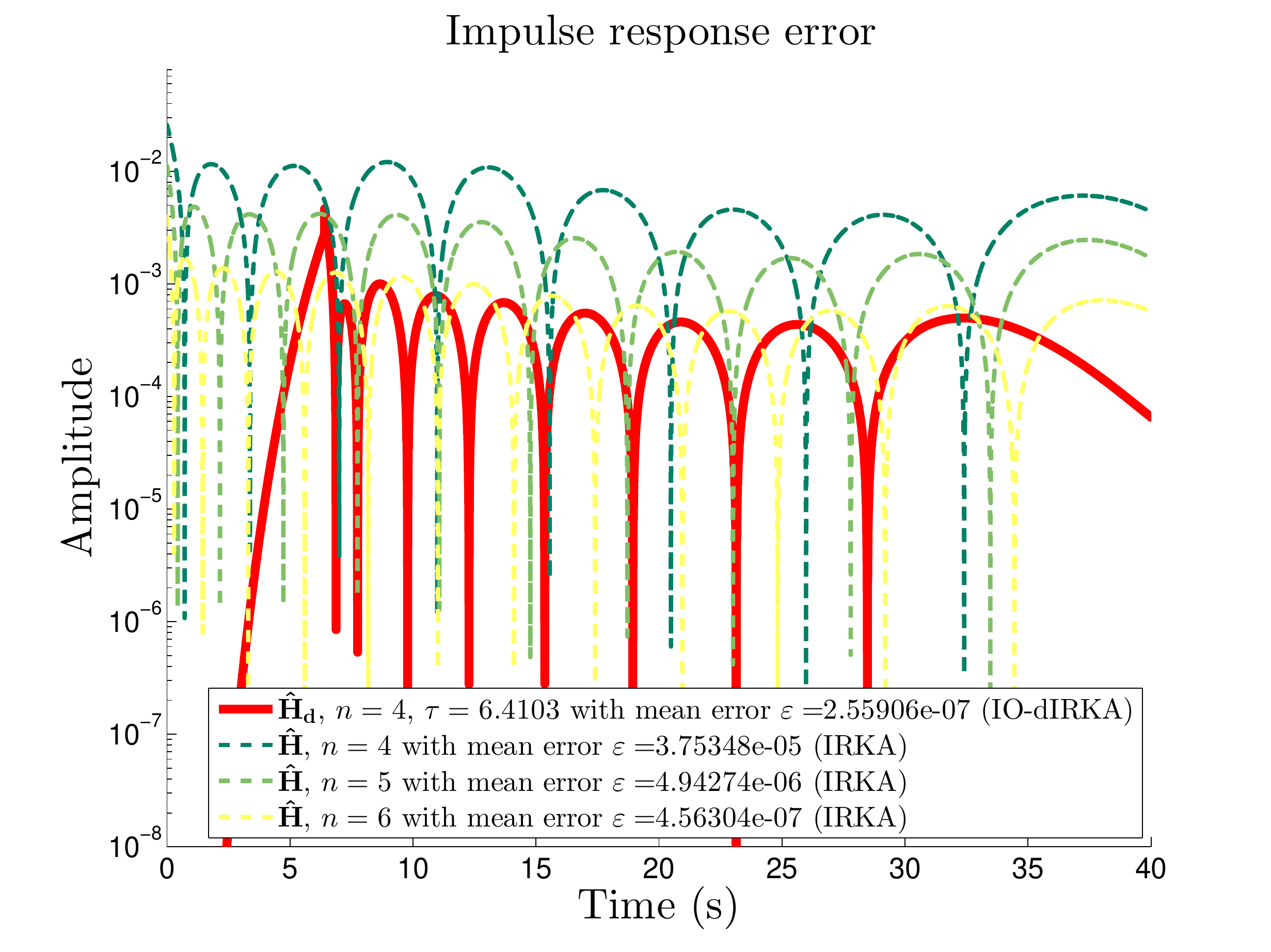}
	 \caption{Impulse response error between the original model $\mathbf{H}$ of order $N=20$ and the input-delay $\Htwo$-optimal model $\hat{\mathbf{H}}_{d}$ of order $n=4$ (solid red line) and the delay-free $\Htwo$-optimal models $\hat{\mathbf{H}}$ of order $n= \{4,5,6\}$ (dashed dark green, light green and yellow lines).}
	 \label{fig:Exemple2r4err}
\end{figure}

With reference to Figure \ref{fig:Exemple2r4}, similar results are obtained in the case of an input delay-dependent approximation of order $n=4$ (using \textbf{IO-dIRKA}) and delay-free approximation of order $n=\{4,5,6\}$ (using \textbf{IRKA}). Then, Figure \ref{fig:Exemple2r4err} shows the impulse response mismatch error for these different configurations. For each reduced order models, the mean square absolute error $\varepsilon$ of the impulse response are computed. The main observation that can be made is that the mismatch error obtained for $\hat{\mathbf{H}}_{d}$ of order $n=4$ is lower that the one obtained by a delay-free model $\hat{\mathbf{H}}$ of order $n=6$ (a better result is obtained for a delay-free model with an order $n=7$). This motivates the use of the specific approximation model delay structure.


\section{Conclusion} \label{sec:conclusion}

The main contribution of this paper is the derivation of the first-order $\mathcal{H}_{2}$ optimality conditions for Problem~\ref{pb:ProblemMIMO}. It forms a direct extension of the bi-tangential interpolation conditions of the delay-free case derived in \cite{gugercin2008h_2, dooren2007}. Theorem~\ref{theo:MIMOoptimH2} establishes that if $\hat{\mathbf{H}}_{d}=\hat{\boldsymbol{\Delta}}_{o}\hat{\mathbf{H}}\hat{\boldsymbol{\Delta}}_{i}$ is a local optimum, then the parameters of this latter verify an extended set of matricial equalities. These ones are of two types: first, \emph{(i)} a subset of interpolation conditions~\eqref{eq:H2interpol} satisfied by the rational part $\hat{\mathbf{H}}$ of $\hat{\mathbf{H}}_{d}$, which generalizes the delay-free case; secondly, \emph{(ii)} a subset of matricial relationships~\eqref{eq:H2tau} focussing on the input/output delay blocks $\hat{\boldsymbol{\Delta}}_{o},~\hat{\boldsymbol{\Delta}}_{i}$. These conditions all are dependent on the reduced order model parametrization described by  $\hat{\mathbf{b}}_{k},~\hat{\mathbf{c}}_{k},~\hat{\lambda}_{k},~\hat{\tau}_{l}$ and $\hat{\gamma}_{m}$, and solving Problem~\ref{pb:ProblemMIMO} requires to tackle a non-convex optmization problem. Nevertheless, an algorithm referred to as \textbf{IO-dIRKA}, has been proposed to practically address this issue. This latter decorrelates the decision variables between them by solving, firstly for given $\hat{\boldsymbol{\Delta}}_{i},~\hat{\boldsymbol{\Delta}}_{o}$ matrices, an optimal $\mathcal{H}_{2}$ approximation problem, and then, in a second stage, a nonlinear maximization problem~\eqref{eq:optDelta} to determine the optimal values of the delays. Both optimizations rely on descent methods, taking benefits from the analytical expressions of the gradients of the $\mathcal{H}_{2}$ mismatch gap $\nabla\mathcal{J}$. Numerical experiment have also been presented, illustrating the benefit of the proposed approximation delay structure with respect to standard delay-free approximation methods.

\bibliographystyle{elsarticle-num} 
\bibliography{igorBiblio}

\end{document}